# Properties of Space MIMO Communication Channels

Richard J. Barton, Member

*Abstract*— This paper discusses the characteristics of a space-to-space multiple-input, multiple-output (MIMO) communication channel that distinguish it from more common terrestrial MIMO communication channels. These characteristics imply that the channel matrices for space communication channels have a particularly simple structure that leads to statistical characteristics that are both predictable and readily controllable using clusters of satellites as distributed communication nodes. Furthermore, the extremely high cost of launching mass into space introduces a constraint into the channel capacity equation that leads to a spectral-efficiency vs. energy-efficiency tradeoff for space MIMO communication that is fundamentally different from the tradeoff that is generally considered applicable to terrestrial MIMO communication systems.

I. INTRODUCTION

In this paper, we consider distributed multiple-input, multiple-output (MIMO) communication in the context of a space-to-space communication system. Space-to-ground communication systems can be expected to have many of the same characteristics, but the distinguishing characteristics of space MIMO channels are most clearly revealed in the space-to-space environment. The two aspects of the space MIMO channel that distinguish it most dramatically from terrestrial applications are extremely long range and a free-space, line-of-sight (LOS) environment that is free of any significant scattering. In particular, we consider communication distances in the range of a few hundred thousand kilomenters (e.g., Earth to Moon) to a few hundred million killometers (e.g., Earth to Mars) and we assume a simple distance-squared power-loss model.

The LOS distributed MIMO problem has been considered in recent years by several authors in the context of cooperative communication in a distributed (terrestrial) wireless network [1-5]. Of particular interest here is a recent paper [5] that rigorously establishes that the number of degrees of freedom on a distributed MIMO channel in a two-dimensional environment is bounded below by a simple relationship between the distribution of the antennas, the distance between the antenna clusters, and the wavelength of the carrier frequency.

Throughout this paper, we are interested in comparing the performance of a conventional single-input, single-output (SISO) space communication system with an "equivalent" $M \times M$ MIMO system. For the SISO system, the channel is characterized by the *channel gain g* and the input *signal-to-noise ratio* (SNR) $\gamma$. These are given by

$$g = \frac{A_T A_R L}{\lambda^2 d^2}, \text{ and } \gamma = \frac{P}{BN_0},$$

respectively, where $B$ is the bandwidth of the transmitted signal, $P$ is the transmitted power, $A_T$ and $A_R$ are the effective areas of the (arbitrarily designated) transmitter and receiver antennas, respectively, $d$ is the range between the two antennas, $\lambda$ is the wavelength at the carrier frequency, $N_0$ is the power spectral density of the AWGN on the baseband equivalent (i.e., complex-valued) channel, and $L$ is a factor that represents the cumulative effect of additional unmodeled losses on the channel such as circuit losses, receiver noise figure, polarization losses, etc. Note that for this paper, since we are dealing with space communication links, we will always assume that $g \ll 1$.

For the equivalent MIMO channel, we also need to know the channel matrix **H**, which is given by

This paragraph of the first footnote will contain the date on which you submitted your paper for review. It will also contain support information, including sponsor and financial support acknowledgment. For example, "This work was supported in part by the U.S. Department of Commerce under Grant BS123456".
R. J. Barton is with NASA Johnson Space Center, Houston, TX 77058 (richard.j.barton@nasa.gov).

$$\mathbf{H} = \begin{bmatrix} h_{11} & h_{12} & \cdots & h_{1M} \\ h_{21} & h_{22} & \cdots & h_{2M} \\ \vdots & \vdots & \cdots & \vdots \\ h_{M1} & h_{M2} & \cdots & h_{MM} \end{bmatrix},$$

where $\{h_{ij}\}$, $i, j = 1, 2, \ldots, M$, represent the complex-valued amplitude and phase couplings between receive antenna *i* and transmit antenna *j*. For a terrestrial system, the equivalent MIMO channel is generally considered to be one in which the total transmit power, distributed across all of the individual transmitting antennas (often uniformly) is also given by *P*, but each of the transmitting and receiving antennas still has area $A_T$ or $A_R$, respectively. That is, the terrestrial MIMO system implicitly includes a receiver *array gain* of *M* and the same total *equivalent isotropic radiated power* (EIRP) *P* as the SISO system.

For the space MIMO channel, if we really want to compare the performance of a MIMO architecture to a SISO architecture in any meaningful way, we must fix not only the total transmitted power but the total antenna aperture area at each end of the link. This is due to the fact that launch costs, which are directly related to mass and hence both system power and system antenna aperture area, are a very large (often dominant) factor in the total cost of deploying and operating a space communication system, and it means that we really need to scale the individual tranmsitter and receiver antennas to have area $A_T/M$ and $A_R/M$, respectively. This implies that there is no receiver array gain associated with a space MIMO system, and the total EIRP from the transmitter array actually *decreases* by a factor of *M* with respect to the SISO system. At first blush, this would seem to imply that a space MIMO system can never be a good idea, and this may be the reason that MIMO antenna systems have not received more attention for space communication systems in the past. However, this is most definitely not the case. In fact, it turns out that there are still both spectral-efficiency and energy-efficiency gains that can be achieved with a space MIMO architecture, albeit different than those ususally associated with a terrestrial MIMO system.

In this paper, we establish upper and lower bounds on the ergodic capacity of a space MIMO channel in two related cases. In the first case, the channel is treated as random due to the fact that a fixed number of uniformly sized antennas are randomly distributed over a fixed spherical volume of space at both ends of the link. In this case, both the size of each individual antenna aperture and the transmitter power at each antenna scale linearly with the number of antennas in order to keep the total aperture area and total power fixed, and a single data stream is assumed transmitted from each antenna. This case is consistent with individual average transmitter power constraints at each antenna.

In the second case, which is non-random, the channel is determined by a spherical region of space over which a fixed number of data streams are transmitted from an arbitrary number of antennas with total fixed aperture size that may be distributed aritrarily over the region. The power is scaled linearly over the data streams rather than the antenna elements, but no assumption is made regarding the number of individual antennas utilized to form the aperture or the size of each individual antenna. Hence, the total aperture area and the total power are still fixed, but the power distribution across the antenna elements and the size of each element are allowed to vary arbitrarily. Individual average transmitter power constraints at each antenna are inconsistent with this model. Obviously, the second case is considerably more general than the first, but the second case provides a great deal of insight into what can be achieved in the first.

The remainder of the paper is organized as follows. In Section II, we discuss the previous work in the literature that is most closely related to the results developed in this paper. In Section III, we review the basic technical background on MIMO and introduce the basic analytical model used for space MIMO channels. The main technical results of the paper are developed in Section IV and discussed in some detail in Section V. Finally, the results developed in the paper are reviewd and summarized in Section VI.

## II. Previous Results

As far as this author knows, the behavior of MIMO communication systems in space, per se, has been studied in only one previous paper [6]. That work established the characteristics and benefits of MIMO for space application under the assumption that the multiplexing gain for a MIMO system in space scales with the number of antennas in exactly the same manner as that of a conventional terrestrial MIMO system operating in a rich scattering environment, with a simple constraint on total aperture area added to the equation. In this paper, we establish rigorously when such an assumption is valid and when it is not, and discuss how that capacity might be achieved in a physically realizable system.

Distributed MIMO communication systems in a terrestrial free-space environment have been studied in several previous papers [1-5], some of which seem to draw contradictory conclusions on the number of degrees of freedom, hence the multiplexing gain, available on a such a channel. The apparent contradiction among these results has been studied in a recent paper [5] by investigating the behavior of the number of degrees of freedom as a function of the area over which the multiple antennas are distributed in a free-space terrestrial environment. In the same paper, lower bounds on the number of degrees of freedom on such terrestrial channels have been obtained.

In a somewhat earlier paper [4], upper bounds on the degrees of freedom in a distributed MIMO system operating in a free-space environment are obtained based on the physical characteristics of electromagnetic propagation between communication nodes distributed within a constrained region. These upper bounds are consistent with both the lower bounds derived in [5] and related information-theoretic scaling laws derived in [7,8] but clearly demonstrate that both spatial constraints and the physics of electromagnetic propagation strongly influence the capacity of information flow between nodes distributed in space. The approach taken in this paper exploits the mathematical structure imposed by the spatial constraints and propagation physics characteristic of a deep-space communication channel to determine bounds on the information capacity of such channels rigorously and fairly precisely as a function of relevant channel parameters such as wavelength, antenna aperture area, range, bandwidth, and transmitter power.

## III. Technical Background

In this section, we briefly review the relevant channel capacity results for SISO and MIMO channels in the context of a space communication link.

### A. SISO Capacity Results

Consider first a conventional space communication system implemented by flying a single satellite with a single transceiver and a single antenna at each end of the link. For such a SISO link over an additive white Gaussian noise (AWGN) channel, the maximum achievable spectral efficiency in bits per second per Hertz (b/s/Hz) is given by [9,10]

$$\xi_1 = \log_2\left(1 + \frac{A_T A_R L P}{\lambda^2 d^2 B N_0}\right) = \log_2(1 + \gamma g) \qquad (1)$$

### B. MIMO Capacity Results

Now consider the situation in which the same communication system is *fractionated* onto $M$ satellites at each end of the link, where each satellite is equipped with a single transceiver with power $P/M$, the satellites at one end all have single antennas with aperture area $A_T/M$, and the satellites at the other end all have single antennas with aperture area $A_R/M$. We assume that the satellites are distributed randomly over approximately spherical regions of space denoted by $\mathcal{V}_T$ and $\mathcal{V}_R$, respectively. This situation is illustrated in Figure 1 below.

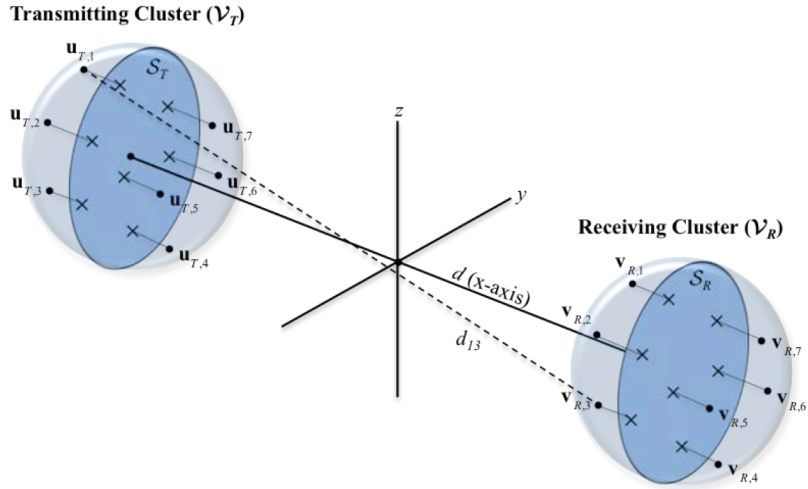

Fig. 1. Space MIMO Example with $M=7$.

Assuming that the volumes of $\mathcal{V}_T$ and $\mathcal{V}_R$ are small compared to the distance $d$ and that the unmodeled losses are the same on each of the point-to-point channels between stations at each end of the link, the channel gain on each individual point-to-point channel will be well approximated by $g/M^2$ and the SNR on each channel will be given by $\gamma/M$. The remaining behavior of the channel is then determined by the structure of the channel matrix $\mathbf{H}$, which will also be random and is discussed in more detail below. Whatever the actual structure of $\mathbf{H}$, the maximum achievable spectral efficiency of the equivalent MIMO channel is then given by [11,12]

$$\xi_M(\mathbf{H}) = \log_2\left(\det\left[\mathbf{I} + \frac{\gamma g}{M^3}\mathbf{H}\mathbf{H}^*\right]\right) = \sum_{i=1}^{M}\log_2\left(1 + \frac{\gamma g}{M^3}|v_i|^2\right), \qquad (2)$$

where $\mathbf{H}^*$ is the complex-conjugate transpose of the matrix $\mathbf{H}$, $\left\{ |\upsilon_1|^2 \geq |\upsilon_2|^2 \geq \cdots \geq |\upsilon_M|^2 \right\}$ are the eigenvalues of $\mathbf{HH}^*$, and $\sum_{i=1}^{M} |\upsilon_i|^2 = M^2$.

Note that the spectral efficiency given by Equation (2) corresponds to channel capacity without channel side information at the transmitter [9]. That is, if the channel model is represented by

$$\mathbf{y} = \sqrt{\frac{g}{M^2}} \mathbf{H}\mathbf{x} + \mathbf{n}, \qquad (3)$$

where $\mathbf{n} \sim \mathcal{N}(\mathbf{0}, \mathbf{I})$ (i.e., complex Gaussian), then (2) represents the maximum achievable spectral efficiency under the constraint $E\{\mathbf{x}\mathbf{x}^*\} = \frac{P}{M}\mathbf{I}$, which can be achieved using a codebook chosen from $\mathbf{x} \sim \mathcal{N}(\mathbf{0}, \frac{P}{M}\mathbf{I})$. We refer to this simply as the *uniform* spectral efficiency. If we assume that full channel side information is available at the transmitter, and we impose only a total transmit power constraint, then the maximum achievable spectral efficiency is given by the so-called *waterfilling solution*, which takes the form [9]

$$\xi_M^{wf}(\mathbf{H}) = \sum_{i=1}^{K} \log_2 \left( 1 + \lambda_i \left[ \frac{\gamma g}{KM^2} + \frac{1}{K}\sum_{j=1}^{K} \frac{1}{\lambda_j} - \frac{1}{\lambda_i} \right] \right), \qquad (4)$$

where $K$ is the largest integer such that

$$\frac{\gamma g}{KM^2} + \frac{1}{K}\sum_{j=1}^{K} \frac{1}{\lambda_j} \geq \frac{1}{\lambda_i},$$

for all $1 \leq i \leq K$. For purposes of this document, we will be concerned exclusively with the properties of $\xi_M(\mathbf{H})$ rather than $\xi_M^{wf}(\mathbf{H})$.

It turns out that for the space MIMO channel over sufficiently large distances $d$, the channel matrix $\mathbf{H}$ has a particularly simple form. To see this, consider the situation illustrated in Figure 1, and let $r_T$ and $r_R$ represent the radii of the spherical volumes $\mathcal{V}_T$ and $\mathcal{V}_R$, respectively. In this case, for sufficiently large $d \gg \max(r_T, r_R)$, the distance $d_{ij}$ between any two transmitter-receiver pairs $\{\mathbf{u}_{T,j}, \mathbf{v}_{R,i}\}$ of the form $\mathbf{u}_{T,j} = (x_{T,j}, y_{T,j}, z_{T,j})^T$ and $\mathbf{v}_{R,i} = (x_{R,i}, y_{R,i}, z_{R,i})^T$ is well approximated by

$$d_{ij} \approx (x_{R,i} - x_{T,j}) + \frac{(y_{R,i} - y_{T,j})^2 + (z_{R,i} - z_{T,j})^2}{2d}.$$

Hence, the elements of $\mathbf{H}$ take the form

$$h_{ij} = e^{-i2\pi \frac{d_{ij}}{\lambda}} \approx e^{-i\frac{2\pi}{\lambda}\left[(x_{R,i}-x_{T,j}) + \frac{(y_{R,i}-y_{T,j})^2+(z_{R,i}-z_{T,j})^2}{2d}\right]} = e^{-i\frac{2\pi}{\lambda}(x_{R,i}-x_{T,j})} e^{-i\frac{\pi}{\lambda d}\left[(y_{R,i}^2+z_{R,i}^2)+(y_{T,j}^2+z_{T,j}^2)\right]} e^{i2\pi \frac{(y_{R,i}y_{T,j})+(z_{R,i}z_{T,j})}{\lambda d}},$$

and $\mathbf{H}$ can be rewritten as

$$\mathbf{H} = (\mathbf{h}_R \mathbf{h}_T^*) \circ \tilde{\mathbf{H}}, \qquad (5)$$

where the notation $\mathbf{A} \circ \mathbf{B}$ denotes the Hadamard (i.e., element-wise) product of the matrices $\mathbf{A}$ and $\mathbf{B}$ and

$$\mathbf{h}_T = \left( e^{i\frac{2\pi}{\lambda}\left(x_{T,1}-\frac{1}{2d}y_{T,1}^2-\frac{1}{2d}z_{T,1}^2\right)}, e^{i\frac{2\pi}{\lambda}\left(x_{T,2}-\frac{1}{2d}y_{T,2}^2-\frac{1}{2d}z_{T,2}^2\right)}, \ldots, e^{i\frac{2\pi}{\lambda}\left(x_{T,M}-\frac{1}{2d}y_{T,M}^2-\frac{1}{2d}z_{T,M}^2\right)} \right)^*,$$

$$\mathbf{h}_R = \left( e^{i\frac{2\pi}{\lambda}\left(x_{R,1}+\frac{1}{2d}y_{R,1}^2+\frac{1}{2d}z_{R,1}^2\right)}, e^{i\frac{2\pi}{\lambda}\left(x_{R,2}+\frac{1}{2d}y_{R,2}^2+\frac{1}{2d}z_{R,2}^2\right)}, \ldots, e^{i\frac{2\pi}{\lambda}\left(x_{R,M}+\frac{1}{2d}y_{R,M}^2+\frac{1}{2d}z_{R,M}^2\right)} \right)^*,$$

$$\tilde{\mathbf{H}} = \left(\tilde{h}_{ij}\right) = \left( e^{i2\pi\frac{\left(y_{R,i}y_{T,j}\right)+\left(z_{R,i}z_{T,j}\right)}{\lambda d}} \right).$$

Note that the matrix $\tilde{\mathbf{H}}$ depends only on the coordinates of the transmitter-receiver pairs $\{\mathbf{u}_{T,j}, \mathbf{v}_{R,i}\}$ projected onto the circular regions in the y-z plane of the local coordinate system denoted by $\mathcal{S}_T$ and $\mathcal{S}_R$, as represented by the dark blue circles in Figure 1. We have the following useful lemma regarding $\mathbf{H}$ and $\tilde{\mathbf{H}}$.

**Lemma 1.** The singular values of $\mathbf{H}$ are equivalent to the singular values of $\tilde{\mathbf{H}}$.

**Proof.** Let $\tilde{\mathbf{H}} = \mathbf{U}\boldsymbol{\Lambda}\mathbf{V}^*$ represent the singular value decomposition (SVD) of the matrix $\tilde{\mathbf{H}}$. It follows from (5) that

$$\mathbf{H} = \mathrm{diag}(\mathbf{h}_R)\tilde{\mathbf{H}}\mathrm{diag}(\overline{\mathbf{h}}_T) = \mathrm{diag}(\mathbf{h}_R)\mathbf{U}\boldsymbol{\Lambda}\mathbf{V}^*\mathrm{diag}(\overline{\mathbf{h}}_T) = \mathbf{W}\boldsymbol{\Lambda}\mathbf{Z}^*,$$

where an overbar indicates complex conjugation

$$\mathbf{W} = \mathrm{diag}(\mathbf{h}_R)\mathbf{U},$$
$$\mathbf{Z} = \mathrm{diag}(\mathbf{h}_T)\mathbf{V},$$

and

$$\mathbf{W}^*\mathbf{W} = \mathbf{U}^*\mathrm{diag}(\overline{\mathbf{h}}_R)\mathrm{diag}(\mathbf{h}_R)\mathbf{U} = \mathbf{U}^*\mathbf{U} = \mathbf{I},$$
$$\mathbf{Z}^*\mathbf{Z} = \mathbf{V}^*\mathrm{diag}(\overline{\mathbf{h}}_T)\mathrm{diag}(\mathbf{h}_T)\mathbf{V} = \mathbf{V}^*\mathbf{V} = \mathbf{I}.$$

Hence, $\mathbf{H} = \mathbf{W}\boldsymbol{\Lambda}\mathbf{Z}^*$ represents a version of the SVD of $\mathbf{H}$, and the singular values of $\mathbf{H}$ are equivalent to the singular values of $\tilde{\mathbf{H}}$. ∎

It follows that to determine $\xi_M$ for the problem illustrated in Figure 1, we can really solve a completely equivalent two-dimensional problem in which the matrix $\mathbf{H}$ is replaced by $\tilde{\mathbf{H}}$. For this equivalent problem, the channel model is now represented by

$$\tilde{\mathbf{y}} = \sqrt{\frac{g}{M^2}}\tilde{\mathbf{H}}\tilde{\mathbf{x}} + \tilde{\mathbf{n}}, \tag{6}$$

and the capacity expression given by Equation (3) remains exactly the same for the two-dimensional problem as the original three-dimensional problem with $\mathbf{H}$ replaced by $\tilde{\mathbf{H}}$. Just to clarify that we are now solving an equivalent but not identical problem, we adopt the notation $\tilde{\xi}_M(\tilde{\mathbf{H}})$ to identify the spectral efficiency for Problem (6), and we note that $\tilde{\xi}_M(\tilde{\mathbf{H}}) = \xi_M(\mathbf{H})$. For the most part, we will refer only to $\tilde{\xi}_M(\tilde{\mathbf{H}})$ throughout the remainder of this paper.

Interestingly, as we discuss in more detail below, Equation (6) implies that the coupling associated with the distributed transmitter and receiver antennas in the original space MIMO problem is very accurately represented by a two-dimensional spatial Fourier transform, in which the spatial signal is associated with the transmitter, and the corresponding spatial-frequency signal (measured in radians of azimuth and elevation) is associated with the receiver. The equivalent two-dimensional problem is illustrated in Figure 2. Note that the three-dimensional transmitter and receiver coordinates $\mathbf{u}_{T,j}$ and $\mathbf{v}_{R,i}$ have been replaced by the two-dimensional equivalents $\tilde{\mathbf{u}}_{T,j} = (y_{T,j}, z_{T,j})^T$ and $\tilde{\mathbf{v}}_{R,i} = (y_{R,i}, z_{R,i})^T$.

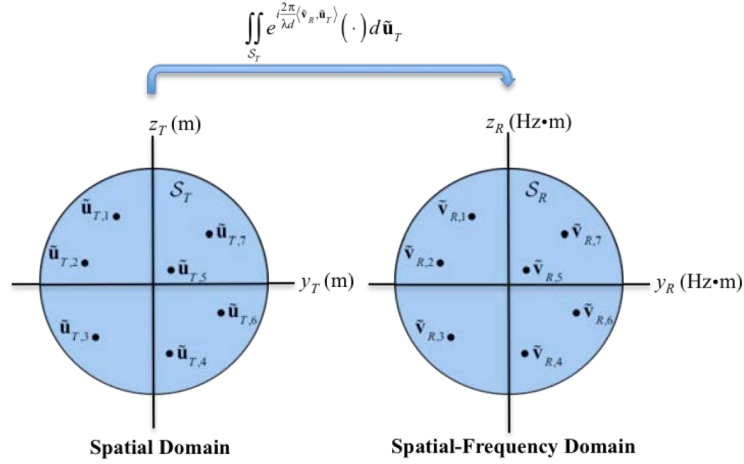

Figure 2. Equivalent 2-D spatial Fourier transform representation.

Making the approximation of internode distances that leads to the conceptual dimensionality reduction for the space MIMO problem can be exploited to simplify the development of both upper and lower bounds for $\tilde{\xi}_M(\tilde{\mathbf{H}})$. A particularly nice result along these lines was provided recently in [5], and that work is referenced in the lemmas proven in the next section. Furthermore, the matrix $\tilde{\mathbf{H}}$ is really just a sampled version of a two-dimensional spatial Fourier transform operator and is intimately connected to the two-dimensional projection operator associated with the well-known space of essentially time- and band-limited functions in two dimensions. This space is spanned by the two-dimensional prolate spheroidal wave functions, and that relationship can be exploited to derive a true capacity expression for the space MIMO communication problem if the definition of "antenna" is relaxed somewhat. Finally, the true capacity can be approached asymptotically by using distributed arrays of physically-realizable antennas operating essentially as beamformers, and the question of interest then becomes how close to the true capacity one can get using feasible antenna arrays and how much training information must be exchanged between the transmitters and receivers.

IV. MAIN RESULTS

In this section, we develop the following results regarding the capacity of a space MIMO channel:

1. We derive both upper and lower bounds for the expected value of $\tilde{\xi}_M(\tilde{\mathbf{H}})$. It will be seen that the upper bound is actually uniform over all realizations of $\tilde{\xi}_M(\tilde{\mathbf{H}})$ rather than just being valid for the expected value, and the lower bound is based on the assumption that the individual antenna nodes are independently and uniformly distributed in the region $\mathcal{S}$. These bounds by themselves establish that the utility of distributed MIMO communication in space depends critically on the relationship between the receiver SNR, the number of distributed antennas at both ends of the link, and the size of the region over which the antennas are distributed.

2. We show that if distributed antennas with radiation patterns matching the two-dimensional prolate spheroidal wave functions are allowed, then the uniform spectral efficiency for the space MIMO channel can be regarded as no longer random. We derive an expression of this uniform spectral efficiency and show that this spectral efficiency can be approached asymptotically by using physically-realizable distributed antenna arrays of sufficient size. We also show that in this case, there is essentially no need for training data to be exchanged between the transmitter and receiver in order to achieve optimal performance.

3. Finally, we derive an expression for what can be considered the true ergodic capacity for the space distributed-antenna MIMO channel by maximizing the deterministic expression for uniform spectral efficiency discussed above over all values of $M \geq 1$.

These results are proven in the collection of lemmas given in this section. Throughout the remainder of the paper, we assume (without loss of much generality for the problem of interest) that $\mathcal{S} = \mathcal{S}_T = \mathcal{S}_R$ is a circle with radius $R$ and area $|\mathcal{S}| = \pi R^2$, and that $|\mathcal{S}|/\lambda d \geq 1$. We will also need to extend our channel model slightly to derive the remaining results. Towards that end, notice that the channel model given by Equation (6) is really just a constrained version of the more general channel model

$$y(\mathbf{v}) = \int_{\mathcal{S}} H(\mathbf{v},\mathbf{u}) x(\mathbf{u}) d\mathbf{u} + n(\mathbf{v}), \tag{7}$$

where $\mathbf{u}, \mathbf{v} \in \mathcal{S}$, $n(\mathbf{v})$ is a complex Gaussian white noise process on $\mathcal{S}$ with power spectral density $N_0$,

$$H(\mathbf{v},\mathbf{u}) = \sqrt{\frac{L}{\lambda^2 d^2}} e^{i\frac{2\pi}{\lambda d}\langle \mathbf{v},\mathbf{u}\rangle},$$

and

$$x(\mathbf{u}) = \sum_{j=1}^{M} x_j f_j(\mathbf{u}),$$

represents symbols $\{x_j\}_{j=1}^{M}$ radiated from $M$ different arbitrarly defined *admissible distributed antennas* with transfer functions $\{f_j(\mathbf{u})\}_{j=1}^{M}$ that satisfy the following properties:

$$\int_{\mathcal{S}} f_i(\mathbf{u}) \overline{f}_j(\mathbf{u}) d\mathbf{u} = \delta_{ij}, \quad i,j = 1,2,\ldots, M,$$

for each $i = 1, 2, \ldots, M$, there exists a set $\mathcal{A}_i \in \mathcal{S}$ such that

$$\mathcal{A}_i = \{\mathbf{u} \in \mathcal{S} : f_i(\mathbf{u}) \neq 0\}, \quad \mathcal{A} = \bigcup_{i=1}^{M} \mathcal{A}_i,$$

and

$$|\mathcal{A}| = \int_{\mathcal{S}} I_{\mathcal{A}}(\mathbf{u}) d\mathbf{u} = A_T, \text{ or } |\mathcal{A}| = \int_{\mathcal{S}} I_{\mathcal{A}}(\mathbf{u}) d\mathbf{u} = A_R,$$

depending on whether the distributed antennas are associated with the transmitting or receiving end of the link, respectively.

That is, in the more general case, the transmitted symbols may be radiated with varying intensity and phase from a continuum of points in $\mathcal{S}$ with aperture area $A_T$. The intensity and phase of the radiation are defined by the transfer functions of the distributed antennas, and the received signal $y(\mathbf{v})$ may be observed (through a set of distributed antennas) over a continuum of points in $\mathcal{S}$ with aperture area $A_R$. Clearly, an upper bound for the maximum achievable uniform spectral efficiency corresponding to Equation (7), which we denote by $\xi_M(\mathcal{S})$, will also be an upper bound for $\tilde{\xi}_M(\tilde{\mathbf{H}})$ corresponding to Equation (6) (hence for $\xi_M(\mathbf{H})$ corresponding to Equation (3) as well).

To make use of this model, we note that the kernel $H(\mathbf{v},\mathbf{u}): \mathcal{S} \rightarrow \mathcal{S}$ defines a compact self-adjoint operator $\mathcal{H}$ mapping the Hilbert space $\mathcal{L}^2(\mathcal{S})$ of square-integrable functions on $\mathcal{S}$ into itself. As such it can be represented as

$$H(\mathbf{v},\mathbf{u}) = \sum_{n=1}^{\infty} v_n p_n(\mathbf{v}) \overline{p}_n(\mathbf{u}),$$

where $\{|v_1|, |v_2|, |v_3|, \ldots\}$ represents the sequence of eigenvalues for the operator $\mathcal{H}$ satisfying

$$|v_1| \geq |v_2| \geq |v_3| \geq \ldots,$$

$$\|\mathcal{H}\|^2 = \sum_{n=1}^{\infty} |v_n|^2 = \int_{\mathcal{S}}\int_{\mathcal{S}} |H(\mathbf{v},\mathbf{u})|^2 d\mathbf{u} d\mathbf{v} = \frac{L}{\lambda^2 d^2}|\mathcal{S}|^2, \qquad (8)$$

and $\{p_n\}_{n=1}^{\infty}$ is a sequence of orthonormal functions that represent the eigenfunctions for $\mathcal{H}$. Furthermore, the set $\{p_n\}_{n=1}^{\infty}$ spans $\mathcal{L}^2(\mathcal{S})$ and either finitely many of the $\{v_n\}_{n=1}^{\infty}$ are nonzero or $|v_n| \rightarrow 0$ as $n \rightarrow \infty$ [13].

Similarly, the restriction of the operator $\mathcal{H}$ to arbitrary admissible antenna aperture sets $\mathcal{A}_T$ and $\mathcal{A}_R$, which we denote by $\mathcal{H}_{\mathcal{A}_T:\mathcal{A}_R}$, is a compact operator defined by the kernel $H_{\mathcal{A}_T:\mathcal{A}_R}(\mathbf{v},\mathbf{u}):\mathcal{A}_T \to \mathcal{A}_R$ given by

$$H_{\mathcal{A}_T:\mathcal{A}_R}(\mathbf{v},\mathbf{u}) = \begin{cases} H(\mathbf{v},\mathbf{u}), & \mathbf{u}\in\mathcal{A}_T, \mathbf{v}\in\mathcal{A}_R, \\ 0, & \text{otherwise.} \end{cases}$$

that maps $\mathcal{L}^2(\mathcal{A}_T)$ into $\mathcal{L}^2(\mathcal{A}_R)$. Formally, we have $\mathcal{H}_{\mathcal{A}_T:\mathcal{A}_R} = \mathcal{P}_{\mathcal{A}_R}\mathcal{H}\mathcal{P}_{\mathcal{A}_T}$, where $\mathcal{P}_{\mathcal{A}_T}$ and $\mathcal{P}_{\mathcal{A}_R}$ are the projection operators onto the spaces $\mathcal{L}^2(\mathcal{A}_T)$ and $\mathcal{L}^2(\mathcal{A}_R)$, respectively.

The properties of $\mathcal{H}_{\mathcal{A}_T:\mathcal{A}_R}$ are very similar to those for $\mathcal{H}$. That is [14],

$$H_{\mathcal{A}_T:\mathcal{A}_R}(\mathbf{v},\mathbf{u}) = \sum_{n=1}^{\infty} \eta_n r_n(\mathbf{v})\overline{q}_n(\mathbf{u}), \tag{9}$$

where $\{|\eta_1|, |\eta_2|, |\eta_3|, \ldots\}$ represents the sequence of singular values for the operator $\mathcal{H}_{\mathcal{A}_T:\mathcal{A}_R}$ satisfying

$$|\eta_1| \ge |\eta_2| \ge |\eta_3| \ge \ldots,$$

and

$$\begin{aligned}\|\mathcal{H}_{\mathcal{A}_T:\mathcal{A}_R}\|^2 &= \sum_{n=1}^{\infty} |\eta_n|^2 \\ &= \int_\mathcal{S}\int_\mathcal{S} |H_{\mathcal{A}_T:\mathcal{A}_R}(\mathbf{u},\mathbf{v})|^2 d\mathbf{u}\,d\mathbf{v} \\ &= \frac{L}{\lambda^2 d^2} A_T A_R.\end{aligned} \tag{10}$$

In this case, $\{q_n\}_{n=1}^{\infty}$ and $\{r_n\}_{n=1}^{\infty}$ are sequences of orthonormal functions that represent the singular functions for $\mathcal{H}_{\mathcal{A}_T:\mathcal{A}_R}$, and the sets $\{q_n\}_{n=1}^{\infty}$ and $\{r_n\}_{n=1}^{\infty}$ span $\mathcal{L}^2(\mathcal{A}_T)$ and $\mathcal{L}^2(\mathcal{A}_R)$, respectively.

For the operator $\mathcal{H}$ considered here, which is really just a two-dimensional spatial Fourier transform operator, the eigenfunctions $\{p_n\}_{n=1}^{\infty}$ are versions of the two-dimensional prolate spheroidal wave functions, which, along with the eigenvalues $\{\nu_n\}_{n=1}^{\infty}$, satisfy the following properties [15,16]:

1. $|\nu_1|^2 \approx |\nu_2|^2 \approx \cdots \approx |\nu_\mathcal{M}|^2 \approx L < 1$ and $\sum_{m=\mathcal{M}+1}^{\infty} |\nu_m|^2 \approx 0$, where $\mathcal{M} = \left\lceil |\mathcal{S}|^2/\lambda^2 d^2 \right\rceil \ge 1$,

2. $\{p_n\}_{n=1}^{\mathcal{M}}$ are essentially space- and band-limited to the set $\mathcal{S}$ in both domains; that is, $p_n(\mathbf{u}) = 0$ for $\mathbf{u} \notin \mathcal{S}$, and $(\mathcal{H}p_n)(\mathbf{v}) \approx 0$ for $\mathbf{v} \notin \mathcal{S}$.

For the sake of completeness, the exact form of $\{\nu_n\}_{n=1}^{\infty}$ and $\{p_n\}_{n=1}^{\infty}$ are given in Appendix I.

Note that for the restricted operator $\mathcal{H}_{\mathcal{A}_T:\mathcal{A}_R}$, we will generally have $(A_T A_R)/(\lambda^2 d^2) \ll 1$, and the properties of the singular values and singular functions no longer correspond to a space of essentially space- and band-limited functions. However, it is still the case that $\sum_{m=\mathcal{M}+1}^{\infty} |\eta_m|^2 \approx 0$. That is, regardless of the choice of admissible antenna aperture sets $\mathcal{A}_T$ and $\mathcal{A}_R$, the restricted operator $\mathcal{H}_{\mathcal{A}_T:\mathcal{A}_R}$ will still have no more than $\mathcal{M}$ significant singular values.

We formally define the space of *essentially space- and band-limited functions* as $\mathcal{B}(\mathcal{S}) \subset \mathcal{L}^2(\mathcal{S})$, where

$$\mathcal{B}(\mathcal{S}) = \left\{ f \in \mathcal{L}^2(\mathcal{S}) : f = \sum_{n=1}^{\infty} f_n p_n, \ \|f\|_{\mathcal{L}^2(\mathcal{S})}^2 = \sum_{n=1}^{\infty} |f_n|^2 < \infty, \text{ and } \|f\|_{\mathcal{B}(\mathcal{S})}^2 = \sum_{n=1}^{\infty} \frac{1}{|v_n|^2} |f_n|^2 < \infty \right\}.$$

Note that these properties imply that the space $\mathcal{B}(\mathcal{S})$ is dense in $\mathcal{L}^2(\mathcal{S})$, and that the "dimension" of $\mathcal{B}(\mathcal{S})$ is essentially $\mathcal{M} = \left\lceil |\mathcal{S}|^2 / \lambda^2 d^2 \right\rceil$, since if $\|f\|_{\mathcal{B}(\mathcal{S})}^2 = 1$, we must have $|f_n| \leq |v_n| \approx 0$ for all $n > \mathcal{M}$. Furthermore, any components of a function in $\mathcal{L}^2(\mathcal{S})$ that are outside of $\text{sp}\{p_n\}_{n=1}^{\mathcal{M}}$ will be strongly attenuated by $\mathcal{H}$. Also, it is sometimes useful to note that $f_n \to f$ in $\mathcal{B}(\mathcal{S})$ implies that $f_n \to f$ in $\mathcal{L}^2(\mathcal{S})$ since

$$\|f_n - f\|_{\mathcal{L}^2(\mathcal{S})}^2 = \sum_{m=1}^{\infty} |f_{nm} - f_m|^2 \leq \sum_{m=1}^{\infty} \frac{1}{|v_n|^2} |f_{nm} - f_m|^2 = \|f_n - f\|_{\mathcal{B}(\mathcal{S})}^2,$$

and that pointwise convergence is implied as well. The pointwise convergence follows from the fact that the space $\mathcal{B}(\mathcal{S})$ is a *reproducing kernel Hilbert space* [17] with reproducing kernel $K$ given by

$$K(\mathbf{v}, \mathbf{u}) = H^* H(\mathbf{v}, \mathbf{u}) = \int_{\mathcal{S}} H^*(\mathbf{v}, \mathbf{w}) H(\mathbf{w}, \mathbf{u}) d\mathbf{w} = \int_{\mathcal{S}} \overline{H(\mathbf{w}, \mathbf{v})} H(\mathbf{w}, \mathbf{u}) d\mathbf{w}.$$

For the remainder of this paper, we will make the simplifying assumption that $v_m \equiv 0$ for all $m > \mathcal{M} = \left\lceil |\mathcal{S}|^2 / \lambda^2 d^2 \right\rceil$. This amounts to replacing the the operator $\mathcal{H}$ in Equation (7) with a finite-dimensional approximation and studying the capacity of the resulting approximate channel model. When $\left\lceil |\mathcal{S}|^2 / \lambda^2 d^2 \right\rceil \gg 1$, this is in fact a very good approximation. In general, the approximation can be made arbitrarily good by arbitrarily choosing the dimension $\mathcal{M} > \left\lceil |\mathcal{S}|^2 / \lambda^2 d^2 \right\rceil$, but that is of very little interest for revealing the behavior of real space channels. Making this assumption also implies that $\eta_m \equiv 0$ for $m > \mathcal{M}$ for arbitrary arbitrary admissible antenna aperture sets, which is very easy to establish as we discuss below.

Expanding the functions $\{f_i(\mathbf{u})\}_{i=1}^{\mathcal{M}}$, $n(\mathbf{u})$, and $y(\mathbf{v})$ with respect to the basis $\{p_n\}_{n=1}^{\infty}$, we can rewrite Equation (7) as

$$\mathbf{y} = \mathbf{VFx} + \mathbf{n}, \tag{11}$$

where

$$\mathbf{y} = (y_1, y_2, \ldots, y_{\mathcal{M}}, \ldots)^T,$$
$$\mathbf{n} = (n_1, n_2, \ldots, n_{\mathcal{M}}, \ldots)^T,$$
$$\mathbf{x} = (x_1, x_2, \ldots, x_{\mathcal{M}})^T,$$
$$\mathbf{V} = \text{diag}\{v_1, v_2, \ldots, v_{\mathcal{M}}, 0, 0, \ldots\},$$
$$\mathbf{F} = \begin{bmatrix} f_{11} & f_{12} & \cdots & f_{1\mathcal{M}} \\ f_{21} & f_{22} & \cdots & f_{2\mathcal{M}} \\ \vdots & \vdots & \cdots & \vdots \\ f_{\mathcal{M}1} & f_{\mathcal{M}2} & \cdots & f_{\mathcal{M}\mathcal{M}} \\ \vdots & \vdots & \cdots & \vdots \end{bmatrix}.$$

To introduce the constraints on antenna aperture size, which are not explicitly reflected in Equation (11), we need to rewrite the equation in terms of the eigenfunctions $\{q_n\}_{n=1}^{\infty}$ and $\{r_n\}_{n=1}^{\infty}$ associated with arbitrary admissible aperture sets $\mathcal{A}_T$ and $\mathcal{A}_R$. Toward this end, we first expand these eigenfunctions in terms of the basis $\{p_n\}_{n=1}^{\infty}$ of $\mathcal{L}^2(\mathcal{S})$ to get

$$q_n(u) = \sum_{i=1}^{\infty} q_{in} p_i(u), \quad \|\mathbf{q}_n\|^2 = \sum_{i=1}^{\infty} |q_{in}|^2 = 1, \quad \mathbf{q}_n = (q_{1n}, q_{2n}, \ldots)^T,$$

$$r_n(u) = \sum_{i=1}^{\infty} r_{in} p_i(u), \quad \|\mathbf{r}_n\|^2 = \sum_{i=1}^{\infty} |r_{in}|^2 = 1, \quad \mathbf{r}_n = (r_{1n}, r_{2n}, \ldots)^T.$$

If we let $\mathbf{Q} = [\mathbf{q}_1 | \mathbf{q}_2 | \cdots | \mathbf{q}_\mathcal{M} | \cdots]$ and $\mathbf{R} = [\mathbf{r}_1 | \mathbf{r}_2 | \cdots | \mathbf{r}_\mathcal{M} | \cdots]$, then in matrix notation corresponding to the basis expansion with respect to $\{p_n\}_{n=1}^{\infty}$, we have $\mathcal{P}_{\mathcal{A}_T} = \mathbf{QQ}^*$ and $\mathcal{P}_{\mathcal{A}_R} = \mathbf{RR}^*$. Since the columns of $\mathbf{F}$ are already assumed to represent admissible antennas with apertures contained in $\mathcal{A}_T$, we must have $\mathbf{F} = \mathbf{QQ}^*\mathbf{F}$. Furthermore, since the matrix $\mathbf{V}$ corresponds to the operator $\mathcal{H}$ in matrix notation, we can write $\mathcal{H} = \mathbf{V}$ and $\mathcal{H}_{\mathcal{A}_T : \mathcal{A}_R} = \mathbf{RR}^*\mathbf{VQQ}^*$. So projecting the output $\mathbf{y}$ in Equation (11) onto $\mathcal{L}^2(\mathcal{A}_R)$ represented with respect to the basis $\{r_n\}_{n=1}^{\infty}$, Equation (11) becomes

$$\mathbf{v} = \mathbf{R}^*\mathbf{y} = \mathbf{R}^*\mathbf{VQQ}^*\mathbf{Fx} + \mathbf{R}^*\mathbf{n} = (\mathbf{R}^*\mathbf{VQ})\mathbf{Q}^*\mathbf{Fx} + \mathbf{n}_{\mathcal{A}_R},$$

where $\mathbf{n}_{\mathcal{A}_R} = \mathbf{R}^*\mathbf{n}$ represents the AWGN process projected onto $\mathcal{L}^2(\mathcal{A}_R)$ and represented with respect to $\{r_n\}_{n=1}^{\infty}$.

To simplify this equation just a bit further, we note that Equation (9) immediately implies that

$$\mathbf{R}^*\mathbf{VQ} = \mathrm{diag}\{\eta_1, \eta_2, \ldots, \eta_\mathcal{M}, \ldots\} \triangleq \mathbf{H}.$$

Since the matrix $\mathbf{R}^*\mathbf{VQ}$ on the left-hand side of this equation has rank at most equal to $\mathcal{M}$, the right-hand side also has rank at most $\mathcal{M}$ and we must have $\eta_m \equiv 0$ for $m > \mathcal{M}$, as stated previously. Hence, for an arbitrary selection of $M$ admissible antennas corresponding to the matrix $\mathbf{F}$, the channel model of interest becomes

$$\mathbf{v} = \mathbf{H}(\mathbf{Q}^*\mathbf{F})\mathbf{x} + \mathbf{n}_{\mathcal{A}_R} = \mathbf{H}\tilde{\mathbf{F}}\mathbf{x} + \mathbf{n}_{\mathcal{A}_R}, \quad (12)$$

where

$$\tilde{\mathbf{F}} = \mathbf{Q}^*\mathbf{F} = \begin{bmatrix} \tilde{f}_{11} & \tilde{f}_{12} & \cdots & \tilde{f}_{1M} \\ \tilde{f}_{21} & \tilde{f}_{22} & \cdots & \tilde{f}_{2M} \\ \vdots & \vdots & \cdots & \vdots \\ \tilde{f}_{\mathcal{M}1} & \tilde{f}_{\mathcal{M}2} & \cdots & \tilde{f}_{\mathcal{M}M} \\ \vdots & \vdots & \cdots & \vdots \end{bmatrix}$$

is the representation of $\mathbf{F}$ with respect to the basis $\{q_n\}_{n=1}^{\infty}$. It should be noted that $\tilde{\mathbf{F}}$ retains the property that $\tilde{\mathbf{F}}^*\tilde{\mathbf{F}} = \mathbf{I}$ and that the eigenvalues $\{\eta_1, \eta_2, \ldots, \eta_\mathcal{M}, \ldots\}$ of the matrix $\mathbf{H}$ depend upon the choice of the antenna aperture sets $\mathcal{A}_T$ and $\mathcal{A}_R$ but not on the particular choice of the matrix $\tilde{\mathbf{F}}$.

We are now ready to state and prove the main results of this paper.

**Lemma 2.** Let $\tilde{\mathbf{H}} = \mathbf{H}\tilde{\mathbf{F}}$ represent the channel matrix in Equation (12) for an arbitrary choice of admissible antenna aperture sets $\mathcal{A}_T$ and $\mathcal{A}_R$ with an arbitrary choice of admissible antennas represented by $\tilde{\mathbf{F}}$, and let $\xi_M(\mathcal{S}, \tilde{\mathbf{H}})$ represent the uniform spectral efficiency for Problem (12) for that particular choice of $\tilde{\mathbf{H}}$. As before, let $\mathcal{M} = \lceil |\mathcal{S}|^2 / \lambda^2 d^2 \rceil$. Then, for any choice of $\tilde{\mathbf{H}}$, we have

$$\xi_M(\mathcal{S}, \tilde{\mathbf{H}}) \leq \min\{M, \mathcal{M}\} \log_2\left(1 + \frac{\gamma g}{M \cdot \min\{M, \mathcal{M}\}}\right), \quad (13)$$

which in turn immediately implies that

$$\tilde{\xi}_M(\tilde{\mathbf{H}}) \le \min\{M, \mathcal{M}\} \log_2\left(1 + \frac{\gamma g}{M \cdot \min\{M, \mathcal{M}\}}\right), \tag{14}$$

for any $\tilde{\mathbf{H}}$. On the other hand, if the $M$ transmit and receive nodes are independently and uniformly distributed over $\mathcal{S}$, then

$$E\{\tilde{\xi}_M(\tilde{\mathbf{H}})\} \ge \frac{\frac{M}{4}\log_2\left(1 + \frac{\lambda g}{2M^2}\right)}{\left(2 - \frac{1}{M}\right) + \frac{32}{9\pi}\left(M - 2 + \frac{1}{M}\right) \Big/ \frac{|\mathcal{S}|}{\lambda d}}. \tag{15}$$

**Proof.** To prove Equations (13) and (14), note first that it follows directly from Equation (2) and Jensen's inequality that

$$\tilde{\xi}_M(\tilde{\mathbf{H}}) = \sum_{i=1}^{M} \log_2\left(1 + \frac{\gamma g}{M^3}|v_i|^2\right) \le M \log_2\left(1 + \frac{\gamma g}{M^4}\sum_{i=1}^{M}|v_i|^2\right) = M\log_2\left(1 + \frac{\gamma g}{M^2}\right).$$

Also note that the diagonal structure of the matrix $\mathbf{H}$ implies that the achievable data rate in Equation (12) can be always be maximized by letting

$$\tilde{\mathbf{F}} = \begin{bmatrix} \mathbf{I}_M \\ \mathbf{0} \end{bmatrix},$$

where $\mathbf{I}_M$ is the $M \times M$ identify matrix. It then follows from Equations (10), (12), and Jensen's inequality that, for $M \le \mathcal{M}$,

$$\xi_M(\mathcal{S}, \tilde{\mathbf{H}}) = \log_2 \det\left(\mathbf{I} + \frac{\gamma}{M}\tilde{\mathbf{H}}\tilde{\mathbf{H}}^*\right) \le \log_2 \det\left(\mathbf{I} + \frac{\gamma}{M}\mathbf{H}\begin{bmatrix}\mathbf{I}_M\\ \mathbf{0}\end{bmatrix}\begin{bmatrix}\mathbf{I}_M & \mathbf{0}\end{bmatrix}\mathbf{H}^*\right) = \sum_{m=1}^{M}\log_2\left(1 + \frac{\gamma}{M}|\eta_m|^2\right)$$

$$\le M\log_2\left(1 + \frac{\gamma}{M^2}\sum_{m=1}^{M}|\eta_m|^2\right) \le M\log_2\left(1 + \frac{\gamma}{M^2}\sum_{m=1}^{\infty}|\eta_m|^2\right) = M\log_2\left(1 + \frac{\gamma g}{M^2}\right).$$

However, recalling that $\eta_m \equiv 0$ for $m > \mathcal{M}$, we also have that, for $M > \mathcal{M}$,

$$\xi_M(\mathcal{S}, \tilde{\mathbf{H}}) = \log_2 \det\left(\mathbf{I} + \frac{\gamma}{M}\tilde{\mathbf{H}}\tilde{\mathbf{H}}^*\right) \le \log_2 \det\left(\mathbf{I} + \frac{\gamma}{M}\mathbf{H}\begin{bmatrix}\mathbf{I}_M\\ \mathbf{0}\end{bmatrix}\begin{bmatrix}\mathbf{I}_M & \mathbf{0}\end{bmatrix}\mathbf{H}^*\right) = \sum_{m=1}^{\mathcal{M}}\log_2\left(1 + \frac{\gamma}{M}|\eta_m|^2\right)$$

$$\le \mathcal{M}\log_2\left(1 + \frac{\gamma}{M\mathcal{M}}\sum_{m=1}^{\mathcal{M}}|\eta_m|^2\right) = \mathcal{M}\log_2\left(1 + \frac{\gamma g}{M\mathcal{M}}\right).$$

Putting these results together proves Equations (13) and (14). The proof of Equation (15) is a bit more involved, and relies on the arguments used to prove Lemma 2.1 in [5]. Since this proof is rather lengthy, it is relegated to Appendix II. ∎

**Lemma 3**. Let $\xi_M(\mathcal{S}) = \sup_{\tilde{\mathbf{H}}} \xi_M(\mathcal{S}, \tilde{\mathbf{H}})$. Then

$$\xi_M(\mathcal{S}) \ge \sum_{m=1}^{M}\log_2\left(1 + \frac{\gamma}{M}\cdot\frac{A_T A_R}{|\mathcal{S}|^2}|v_m|^2\right) \approx \min\{M, \mathcal{M}\}\log_2\left(1 + \frac{\gamma g}{M\mathcal{M}}\right),$$

and this lower bound on $\xi_M(\mathcal{S})$ can be approached asymptotically using physically realizable collections of individual antenna elements.

**Proof**. Let $\{\mathcal{S}_N\}_{N=1}^{\infty}$ be a sequence of partitions of the set $\mathcal{S}$ such that

$$\mathcal{S}_N = \left\{\mathcal{S}_{N,i} \subset \mathcal{S} \,\Big|\, \mathcal{S} = \bigcup_{i=1}^{K_N}\mathcal{S}_{N,i},\ \mathcal{S}_{N,i}\cap\mathcal{S}_{N,j} = \varnothing, i\ne j,\ |\mathcal{S}_{N,i}| \approx \frac{|\mathcal{S}|}{N}\right\},\quad N = 1,2,\ldots,\infty.$$

Let $\mathcal{B}_\delta(\mathbf{u})$ be a disc of area $\delta$ centered at the origin in $\mathcal{S}$, and let $\{\mathcal{U}_N\}_{N=1}^\infty$ be a corresponding sequence of sets such that

$$\mathcal{U}_N = \left\{ \mathbf{u}_{N,i} \in \mathcal{S} \,\middle|\, \mathcal{B}_{A_T/N}(\mathbf{u} - \mathbf{u}_{N,i}) \subset \mathcal{S}_{N,i},\ \mathcal{B}_{A_R/N}(\mathbf{u} - \mathbf{u}_{N,i}) \subset \mathcal{S}_{N,i},\ i = 1,2,\ldots,K_N \right\},$$

and

$$\int_\mathcal{S} \overline{\hat{p}_{N,m}(\mathbf{v})} \int_\mathcal{S} H(\mathbf{v},\mathbf{u}) \hat{p}_{N,n}(\mathbf{u}) d\mathbf{u}\,d\mathbf{v} = \int_\mathcal{S} \left( \sum_{i=1}^{K_N} \overline{p_m(\mathbf{v}_{N,i})} I_{\mathcal{S}_{N,i}}(\mathbf{v}) \right) \int_\mathcal{S} H(\mathbf{v},\mathbf{u}) \left( \sum_{j=1}^{K_N} p_n(\mathbf{u}_{N,j}) I_{\mathcal{S}_{N,j}}(\mathbf{u}) \right) d\mathbf{u}\,d\mathbf{v}$$

$$\xrightarrow[N\to\infty]{} \int_\mathcal{S} \overline{p_m(\mathbf{v})} \int_\mathcal{S} H(\mathbf{v},\mathbf{u}) p_n(\mathbf{u}) d\mathbf{u}\,d\mathbf{v} = v_n \delta_{mn},$$

uniformly for all $n,m = 1,2,\ldots,M$. Here, $\delta_{mn}$ represents the Kronecker delta function, $I_\mathcal{A}(\mathbf{u})$ represents the indicator function for the set $\mathcal{A}$, and

$$\hat{p}_{N,m}(\mathbf{u}) = \sum_{i=1}^{N} p_m(\mathbf{u}_{N,i}) I_{\mathcal{S}_{N,i}}(\mathbf{u})$$

is a simple-function approximation of $p_m(\mathbf{u})$. If we now let

$$f_{N,m}^T(\mathbf{u}) = \sqrt{\frac{|\mathcal{S}|}{A^T}} \sum_{i=1}^{N} p_m(\mathbf{u}_{N,i}) \mathcal{B}_{A_T/N}(\mathbf{u} - \mathbf{u}_{N,i}),$$

$$g_{N,m}^R(\mathbf{v}) = \sqrt{\frac{|\mathcal{S}|}{A^R}} \sum_{j=1}^{N} p_m(\mathbf{u}_{N,j}) \mathcal{B}_{A_R/N}(\mathbf{v} - \mathbf{v}_{N,j}),$$

for $m = 1,2,\ldots,M$, represent a particular choice of $M$ admissible distributed antennas derived from $\{\hat{p}_{N,m}(\mathbf{u})\}_{m=1}^M$ at the transmitter and receiver, respectively, then it follows immediately that

$$\int_\mathcal{S} \overline{g_{N,m}^R(\mathbf{v})} \int_\mathcal{S} H(\mathbf{v},\mathbf{u}) f_{N,n}^T(\mathbf{u}) d\mathbf{u}\,d\mathbf{v} = \int_\mathcal{S} \left( \sqrt{\frac{|\mathcal{S}|}{A_R}} \sum_{i=1}^{K_N} \overline{p_m(\mathbf{v}_{N,i})} \mathcal{B}_{A_R/N}(\mathbf{v} - \mathbf{v}_{N,i}) \right) \int_\mathcal{S} H(\mathbf{v},\mathbf{u}) \left( \sqrt{\frac{|\mathcal{S}|}{A_T}} \sum_{j=1}^{K_N} p_n(\mathbf{u}_{N,j}) \mathcal{B}_{A_T/N}(\mathbf{u} - \mathbf{u}_{N,i}) \right) d\mathbf{u}\,d\mathbf{v}$$

$$= \sqrt{\frac{A_T A_R}{|\mathcal{S}|^2}} \int_\mathcal{S} \left( \sum_{i=1}^{K_N} \overline{p_m(\mathbf{v}_{N,i})} I_{\mathcal{S}_{N,i}}(\mathbf{v}) \right) \int_\mathcal{S} H(\mathbf{v},\mathbf{u}) \left( \sum_{j=1}^{K_N} p_n(\mathbf{u}_{N,j}) I_{\mathcal{S}_{N,j}}(\mathbf{u}) \right) d\mathbf{u}\,d\mathbf{v}$$

$$\xrightarrow[N\to\infty]{} \sqrt{\frac{A_T A_R}{|\mathcal{S}|^2}} \int_\mathcal{S} \overline{p_m(\mathbf{v})} \int_\mathcal{S} H(\mathbf{v},\mathbf{u}) p_n(\mathbf{u}) d\mathbf{u}\,d\mathbf{v} = \sqrt{\frac{A_T A_R}{|\mathcal{S}|^2}} v_n \delta_{mn},$$

and the convergence is again uniform over all $n,m = 1,2,\ldots,M$. Hence, for this particular choice of admissible antennas at the transmitter and receiver, we have a version of Problem (6) with $M \times M$ channel matrix $\tilde{\mathbf{H}}_N = (\tilde{h}_{mn}^N)$ that satisfies

$$\sqrt{\frac{g}{M^2}} \tilde{h}_{mn}^N = \int_\mathcal{S} \overline{g_{N,m}^R(\mathbf{v})} \int_\mathcal{S} H(\mathbf{v},\mathbf{u}) f_{N,n}^T(\mathbf{u}) d\mathbf{u}\,d\mathbf{v},$$

and it is clear that

$$\tilde{\xi}_M(\tilde{\mathbf{H}}_N) \xrightarrow[N\to\infty]{} \sum_{m=1}^{M} \log_2 \left( 1 + \frac{\gamma}{M} \cdot \frac{A_T A_R}{|\mathcal{S}|^2} |v_m|^2 \right).$$

It follows that for any $\varepsilon > 0$, we can find $N > 0$ such that

$$\tilde{\xi}_M(\tilde{\mathbf{H}}_N) > \sum_{m=1}^{M} \log_2\left(1 + \frac{\gamma}{M} \cdot \frac{A_T A_R}{|\mathcal{S}|^2}|v_m|^2\right) - \varepsilon,$$

and this value of $\tilde{\xi}_M(\tilde{\mathbf{H}}_N)$ is achieved using a physically realizable collection of individual antenna elements. Hence, it follows that

$$\xi_M(\mathcal{S}) \geq \sum_{m=1}^{M} \log_2\left(1 + \frac{\gamma}{M} \cdot \frac{A_T A_R}{|\mathcal{S}|^2}|v_m|^2\right),$$

and that the lower bound can be approached asymptotically using physically realizable collections of individual antenna elements. The final approximation of the lower bound follows from the fact that $\mathcal{M} = \lceil |\mathcal{S}|^2/\lambda^2 d^2 \rceil$ and the approximations $|v_m| \approx L$ for $m \leq \mathcal{M}$ and $|v_m| \approx 0$ for $m > \mathcal{M}$. ∎

**Corollary 1**. It follows from Lemmas 2 and 3 that

$$\xi_M(\mathcal{S}) \approx \min\{M,\mathcal{M}\}\log_2\left(1 + \frac{\gamma g}{M \cdot \min\{M,\mathcal{M}\}}\right).$$

**Proof**. For $M \geq \mathcal{M}$, the approximate version of the lower bound from Lemma 3 together with the upper bound from Lemma 2 imply that

$$\min\{M,\mathcal{M}\}\log_2\left(1 + \frac{\gamma g}{M \cdot \min\{M,\mathcal{M}\}}\right) = \min\{M,\mathcal{M}\}\log_2\left(1 + \frac{\gamma g}{M\mathcal{M}}\right) \leq \xi_M(\mathcal{S}) \leq \min\{M,\mathcal{M}\}\log_2\left(1 + \frac{\gamma g}{M \cdot \min\{M,\mathcal{M}\}}\right),$$

and the result follows. For $M < \mathcal{M}$, we note that to establish the lower bound on $\xi_M(\mathcal{S})$, we are free to restrict the admissible antenna domain to the circle $\mathcal{S}' \subset \mathcal{S}$ with $|\mathcal{S}'| = \sqrt{M}(\lambda d) < |\mathcal{S}|$. It then follows immediately from the proof of the approximate lower bound in Lemma 3 applied to the restricted domain $\mathcal{S}'$ with $\mathcal{M}' = M$ that

$$\xi_M(\mathcal{S}) \geq \min\{M,\mathcal{M}'\}\log_2\left(1 + \frac{\gamma g}{M\mathcal{M}'}\right) = M\log_2\left(1 + \frac{\gamma g}{M^2}\right) = \min\{M,\mathcal{M}\}\log_2\left(1 + \frac{\gamma g}{M \cdot \min\{M,\mathcal{M}\}}\right).$$

Hence, applying the upper bound from Lemma 2, we once again have

$$\min\{M,\mathcal{M}\}\log_2\left(1 + \frac{\gamma g}{M \cdot \min\{M,\mathcal{M}\}}\right) \leq \xi_M(\mathcal{S}) \leq \min\{M,\mathcal{M}\}\log_2\left(1 + \frac{\gamma g}{M \cdot \min\{M,\mathcal{M}\}}\right),$$

and the result follows. ∎

**Lemma 4**. For fixed values of the SNR $\gamma$ and the channel gain $g$, if we define the ergodic capacity of the space distributed-antenna MIMO channel (in bits/sec/Hz) as

$$\xi_{\gamma,g} = \max_M \sup_{\mathcal{S}} \xi_M(\mathcal{S}),$$

then

$$\xi_{\gamma,g} = \max\left\{M_l \log_2\left(1 + \frac{\gamma g}{M_l^2}\right), M_u \log_2\left(1 + \frac{\gamma g}{M_u^2}\right)\right\},$$

where $M_l = \left\lfloor \sqrt{\frac{\gamma g}{3.9125}} \right\rfloor$ and $M_u = \left\lceil \sqrt{\frac{\gamma g}{3.9125}} \right\rceil$. Furthermore

$$\xi_{\gamma,g} \approx 0.8053\sqrt{\gamma g}.$$

**Proof.** As usual, we take $\mathcal{S}$ to be a circle with radius $R$ and $\mathcal{M} \approx |\mathcal{S}|^2/\lambda^2 d^2$. Then it is clear that from Corollary 1 that

$$\sup_{\mathcal{S}} \xi_M(\mathcal{S}) \approx M\log_2\left(1 + \frac{\gamma g}{M^2}\right).$$

For fixed $\gamma$ and g, we define the function

$$\xi_{\gamma,g}(x) = x\log_2\left(1 + \frac{\gamma g}{x^2}\right),$$

and we note that

$$x_{opt} = \arg\max_{x} \xi_{\gamma,g}(x) \approx \sqrt{\frac{\gamma g}{3.9125}},$$

with

$$\max_{x} \xi_{\gamma,g}(x) = \xi_{\gamma,g}(x_{opt}) \approx 0.8053\sqrt{\gamma g}.$$

If we define $M_l = \lfloor x_{opt} \rfloor$ and $M_l = \lceil x_{opt} \rceil$, then we have

$$\xi_{\gamma,g} = \max\{\xi_{\gamma,g}(M_l), \xi_{\gamma,g}(M_u)\} = \max\left\{M_l\log_2\left(1 + \frac{\gamma g}{M_l^2}\right), M_u\log_2\left(1 + \frac{\gamma g}{M_u^2}\right)\right\} \approx 0.8053\sqrt{\gamma g},$$

as claimed. ∎

## V. Discussion

Consider first the implications of Lemma 2. Note that only the lower bound in Lemma 2 involves the expected value of the uniform spectral efficiency that depends on the antennas having a uniform distribution over $\mathcal{S}$. The upper bounds apply to every realization of Problems (6) or (7) and are true regardless of the distribution of the antennas. These bounds establish several interesting characteristics of space MIMO communication. First and foremost among these is that if the antennas are distributed over a sufficiently large region $\mathcal{S}$, then

$$E\{\tilde{\xi}_M(\tilde{\mathbf{H}})\} \sim M\log_2\left(1 + \frac{\gamma g}{M^2}\right).$$

This relationship was exploited in a previous paper [6] to study the optimal number of antennas for a space MIMO communication system as a function of the well-known energy-efficiency, spectral-efficiency tradeoff on the channel. It was shown in that work that for scenarios in which the available $E_b/N_0$ budget at the transmitter is quite low, a conventional SISO space communication link will always provide the highest achievable spectral efficiency on the channel. Conversely, it was shown that as the $E_b/N_0$ available at the transmitter increases, the optimal value of $M$ grows, and that for any chosen level of $E_b/N_0$ there is a unique value of $M$ that maximizes the spectral efficiency. This is distinctly different behavior than one would see for a terrestrial channel in which the total antenna aperture size is not constrained as the number of antennas grows.

Lemma 2 also shows that as the number of antennas in a space MIMO system grows, the antennas must be distributed over larger and larger regions in order to realize any gains from spatial multiplexing. This is a direct result of the fact that there is no scattering on a space link to provide spatial diversity with minimal separation between the antennas, which is not surprising, but it is somewhat surprising that a random distribution of antennas provides sufficient diversity to guarantee good average performance.

The final interesting characteristic of the bounds derived in Lemma 2 is the behavior of the gap between the upper and lower bounds. On the one hand, the upper bound by itself implies that the maximum number of degrees of freedom on the channel is given by $\mathcal{M} \approx |\mathcal{S}|^2/\lambda^2 d^2$. On the other hand, the lower bound by itself implies that performance may begin to degrade for any value of $M > |\mathcal{S}|/\lambda d$, which implies that the maximum number of degrees of freedom may be closer to $|\mathcal{S}|/\lambda d$.

The gap between the upper and lower bounds in Lemma 2 is closed by Corollary 1, which establishes that

$$\xi_M(\mathcal{S}) \approx \min\{M, \mathcal{M}\} \log_2\left(1 + \frac{\gamma g}{M \cdot \min\{M, \mathcal{M}\}}\right).$$

This implies that the maximum number of degrees of freedom on the channel is indeed given by $\mathcal{M} \approx |\mathcal{S}|^2 / \lambda^2 d^2$. Furthermore, the proof of Lemma 3 clearly demonstrates that this value of the uniform spectral efficiency can be approached asymptotically by using physically realizable distributed antenna arrays having a finite number of antenna elements at both ends of the channel. The particular realization of antenna arrays constructed in the proof of Lemma 3 uses the same simple-function approximations of the basis functions $\{p_n\}_{n=1}^{M}$ at both ends of the channel to derive the transmitter and receiver antenna arrays. In practice, doing this would obviously require exchange of sufficient training information between the transmitter and receiver to specify the antenna arrays completely, and this would probably be nearly impossible to achieve given the dynamic nature of the ephemerides and the large time-delay associated with two constellations of satellites in widely separated orbits. Fortunately, the proof works just as well by using completely arbitrary – possibly total different and independently derived – simple-function approximations to derive the distributed antennas at the two ends of the link. This makes the proof somewhat messier in terms of notation, but otherwise changes almost nothing in the argument.

Hence, to achieve the maximum uniform spectral efficieny for the distributed-antenna space MIMO channel for a given number of independent data streams and a given radius of antenna array element distribution at both ends of the link, it is only necessary for the transmitter and receiver to exchange enough training information to establish and maintain a common coordinate system. As long as the transmitter and receiver both know where their own satellites are relative to the common coordinate system, then both can independently construct admissible antenna sets that will be asymptotically good.

The question of how many individual antenna elements (equivalently, communication nodes) must be used to construct a set of suitable distributed antennas is a different issue that is not addressed in Lemma 3. It is clear from the proof of Lemma 3 that all $M \leq \mathcal{M}$ data streams can be transmitted from the same collection of distributed nodes, which together comprise $M$ different distributed antennas by varying the phase and amplitude of radiation across the fixed array of distributed elements. However, the number (say $N$) of nodes needed to achieve near-optimal performance is not addressed in the proof of Lemma 3. To determine the required number $N$ of elements in the distributed antenna array, it becomes necessary to study the smoothness characteristics of the two-dimensional prolate spheroidal wave functions, which is beyond the scope of this work; however, based on the Nyquist spatial sampling rate associated with the space-bandwidth characteristics of those functions, one can conjecture that the total number of required elements is approximately $N \geq |\mathcal{S}|^2 / \lambda^2 d^2$. Using the same logic, one can also conjecture that as long as the $N$ nodes are fairly uniformly distributed throughout $\mathcal{S}$ at both ends of the link, the performance should be reasonably stable. That is, the overall achievable spectral efficiency should not be very sensitive to either the number of antenna nodes or small variations in the relative position of the nodes as long as the node density is above $|\mathcal{S}|/\lambda^2 d^2$ nodes per square meter and the distribution is roughly uniform.

Finally, Lemma 4 demonstrates that the quantity

$$\xi_{\gamma,g} \approx 0.8053\sqrt{\gamma g}$$

can be regarded as the maximum achievable spectral efficiency for a space communication channel characterized by transmitter SNR $\gamma$ and channel gain $g$ as long as $\gamma g \geq 3.9125$. To achieve that spectral efficiency, $M \approx \sqrt{\gamma g / 3.9125}$ independent data streams must be transmitted between two distributed arrays of $N \geq M$ communication nodes distributed over an area of approximately

$$|\mathcal{S}| = \sqrt{M} \cdot \lambda d$$

square meters at both ends of the link.

## VI. Conclusion

In this paper, we have studied the characteristics of a space-based MIMO communication system that distinquish it from more familiar terrestrial MIMO communication systems that operate in richer scattering environments. We have shown that the lack of scattering and the extreme ranges encountered on space-based communication links lead to anayltical channel models with a relatively simple structure that not only simplifies statistical analysis, but also reveals a great deal of geometrical structure that is not present in most terrestrial envionments. We have also argued that the high cost of launching communication infrastructure into space introduces an antenna aperture constraint into the information-theoretic expressions for channel capacity that must be considered in order to produce meaningful results for purposes of performance analysis and system design. This additional constraint together with the additional geometrical structure that is not present in terrestrial environments results in fundamentally different behavior in achievable spectral efficiency for space MIMO channels than for terrestrial MIMO channels.

We have analyzed the channel capacity of space MIMO channels using both a stochastic channel model based on the assumption of random distribution of antenna nodes over fixed spherical regions of space and a more general deterministic channel model corresponding to the function space of admissible distributed antenna solutions satisfying the antenna aperture constraint over the same spherical regions of space. For the stochastic channel model, we have derived fairly precise upper and lower bounds on the expected value of the achievable spectral efficiency, as a function of the number of independent data streams transmitted over the channel, that are distinct from but consistent with previously derived bounds for free-space terrestrial environments. We have shown that there is a gap between the upper and lower bounds that makes it difficult to identify unambiguously the number of degrees of freedom available on a space MIMO channel using only these bounds.

For the more general deterministic channel model, we have also derived precise upper and lower bounds on achievable spectral efficiency and shown that they are identical. We have demonstrated that the resulting unique achievable spectral efficiency can be approached asymptotically using physically realizable arrays of antenna elements distributed over the given regions of space. We have also demonstrated that the optimal spectral efficiency can be achieved with no training data exchanged between the transmitter and receiver ends of the link beyond what is needed for the two ends of the link to identify a common coordinate system. Finally, we have shown that maximizing the expression for achievable spectral efficiency (for a fixed number of independent data streams transmitted between given spherical regions of space) over all possible values for the number of data streams and the volume of the antenna distribution domains yields what can be regarded as the maximum achievable spectral efficiency for a MIMO space communication channel characterized by given transmitter SNR and channel gain.

It is worth noting at this point that we have ignored in this paper the cost of exchanging information between communication nodes at either end of the link. That is, when deriving the expressions for achievable spectral efficiency, we have assumed that the power required to exchange data between the cooperating communication nodes at either end of the link prior to transmission or decoding and the impact of that data exchange on overall data rate are both insignificant. Given the fact that the ratio of the communication range to the radius of required antenna distribution volume for most space applications is several orders of magnitude, at least, this assumption seems justified.

VII. APPENDIX I

Let $\mathcal{D}$ represent the unit circle in $\mathbb{R}^2$, and recall that $\mathcal{S} = R\mathcal{D}$. The eigenvalues and eigenfunctions for Problem (8) are given by

$$v_n = \sqrt{\frac{L}{\lambda^2 d^2}} R^2 \alpha_n,$$

$$p_n(\mathbf{u}) = \psi_n(\mathbf{u}/R), \quad \mathbf{u} \in \mathcal{S},$$

where $\{\alpha_n\}_{n=1}^{\infty}$ and $\{\psi_n(\mathbf{u})\}_{n=1}^{\infty}$ are the eigenvalues and eigenfunctions of the integral equation

$$\alpha_n \psi_n(\mathbf{v}) = \iint_{\mathcal{D}} e^{i\frac{2\pi R^2}{\lambda d}\langle \mathbf{v}, \mathbf{u}\rangle} \psi_n(\mathbf{u}) d\mathbf{u}, \quad \mathbf{v} \in \mathcal{D}.$$

The solutions to this integral equation have been widely studied [14,15] and are given by the *prolate spheroidal wave functions*

$$\psi_{|N|,m}(r,\theta) = R_{|N|,m}(r) e^{i\frac{2\pi}{\lambda d} N\theta}, \quad N = 0, \pm 1, \pm 2, \ldots, \quad m = 0, 1, 2, \ldots,$$

(now doubly indexed and given in terms of the polar coordinates $\mathbf{u} = (r,\theta)$) and the associated eigenvalues

$$\alpha_{N,m} = 2\pi i^N \beta_{|N|,m},$$

where

$$\beta_{|N|,m} R_{|N|,m}(r) = \int_0^1 J_{|N|}\left(\frac{2\pi R^2}{\lambda d} rr'\right) R_{|N|,m}(r') r' dr', \quad 0 \leq r \leq 1,$$

and $\{J_N(r)\}_{N=0}^{\infty}$ are the Bessel functions of the first kind [18].

## VIII. APPENDIX II

The proof of Equation (15) given in this appendix is similar in many respects to the proof of Lemma 2.1 in [5].

**Proof of Equation (15).** Let $\{\upsilon_i\}_{i=1}^{M}$ represent the sequence of singular values of the matrix $\tilde{\mathbf{H}} = \{\tilde{h}_{ij}\}$, as before, and recall that

$$E\{\tilde{\xi}_M(\tilde{\mathbf{H}})\} = E\left\{\log_2\left(\det\left[\mathbf{I} + \frac{\gamma g}{M^3}\tilde{\mathbf{H}}\tilde{\mathbf{H}}^*\right]\right)\right\} = M \cdot E\left\{\frac{1}{M}\sum_{i=1}^{M}\log_2\left(1 + \frac{\gamma g}{M^3}|\upsilon_i|^2\right)\right\} = M \cdot E\left\{\sum_{i=1}^{M}\log_2\left(1 + \frac{\gamma g}{M^3}|\upsilon|^2\right)\right\},$$

where $\upsilon$ represents a randomly selected singular value. Also note that

$$E\{|\upsilon|^2\} = E\left\{\frac{1}{M}\sum_{i=1}^{M}|\upsilon_i|^2\right\} = E\left\{\frac{1}{M}\mathrm{Tr}(\tilde{\mathbf{H}}\tilde{\mathbf{H}}^*)\right\} = M,$$

and

$$E\{|\upsilon|^4\} = E\left\{\frac{1}{M}\mathrm{Tr}(\tilde{\mathbf{H}}\tilde{\mathbf{H}}^*\tilde{\mathbf{H}}\tilde{\mathbf{H}}^*)\right\}.$$

Now, following the proof of Lemma 2.1 in [5], it follows from the Paley-Zigmund inequality [19, (21.12)] that for any $0 \leq t \leq M$, we have

$$E\{\tilde{\xi}_M(\tilde{\mathbf{H}})\} \geq M \cdot \log_2\left(1 + \frac{\gamma g}{M^3}t\right)\frac{(M-t)^2}{E\{|\upsilon|^4\}}.$$

Hence, choosing $t = M/2$, we get

$$E\{\tilde{\xi}_M(\tilde{\mathbf{H}})\} \geq \frac{M^3}{4} \cdot \log_2\left(1 + \frac{\gamma g}{2M^3}\right)\frac{1}{E\{|\upsilon|^4\}},$$

and we are left with the task of finding a suitable upper bound for $E\{|\upsilon|^4\}$. Toward that end, notice that

$$E\{|\upsilon|^4\} = E\left\{\frac{1}{M}\mathrm{Tr}(\tilde{\mathbf{H}}\tilde{\mathbf{H}}^*\tilde{\mathbf{H}}\tilde{\mathbf{H}}^*)\right\} = \frac{1}{M}\sum_{i=1}^{M}\sum_{j=1}^{M}\sum_{k=1}^{M}\sum_{l=1}^{M}E\{\tilde{h}_{ik}\overline{\tilde{h}_{jk}}\tilde{h}_{il}\overline{\tilde{h}_{jl}}\}$$

$$= \frac{1}{M}\sum_{i=1}^{M}\sum_{k=1}^{M}E\{|\tilde{h}_{ik}|^4\} + \frac{1}{M}\sum_{i=1}^{M}\sum_{k=1}^{M}\sum_{\substack{l=1\\l\neq k}}^{M}E\{|\tilde{h}_{ik}|^2|\tilde{h}_{il}|^2\} + \frac{1}{M}\sum_{k=1}^{M}\sum_{i=1}^{M}\sum_{\substack{j=1\\j\neq i}}^{M}E\{|\tilde{h}_{ik}|^2|\tilde{h}_{jk}|^2\} + \frac{1}{M}\sum_{i=1}^{M}\sum_{\substack{j=1\\j\neq i}}^{M}\sum_{k=1}^{M}\sum_{\substack{l=1\\l\neq k}}^{M}E\{\tilde{h}_{ik}\overline{\tilde{h}_{jk}}\tilde{h}_{il}\overline{\tilde{h}_{jl}}\}$$

$$= \frac{M^2}{M} + \frac{2M^2(M-1)}{M} + \frac{M^2(M-1)^2}{M}E\{\tilde{h}_{ik}\overline{\tilde{h}_{jk}}\tilde{h}_{il}\overline{\tilde{h}_{jl}}\}.$$

Assuming that the points $\{\tilde{\mathbf{u}}_{T,i}\}_{i=1}^{M}$ and $\{\tilde{\mathbf{v}}_{T,j}\}_{j=1}^{M}$ are independent and uniformly distributed over the set $\mathcal{S}$, this becomes

$$E\{|\upsilon|^4\} = 2M^2 - M + \frac{M(M-1)^2}{|\mathcal{S}|^4}\int_{\mathcal{S}}\int_{\mathcal{S}}\int_{\mathcal{S}}\int_{\mathcal{S}}\begin{bmatrix}e^{i\frac{2\pi}{\lambda d}\langle\mathbf{u}_{R,i},\mathbf{v}_{T,k}\rangle} \cdot e^{-i\frac{2\pi}{\lambda d}\langle\mathbf{u}_{R,j},\mathbf{v}_{T,k}\rangle}\\ \cdot e^{-i\frac{2\pi}{\lambda d}\langle\mathbf{u}_{R,i},\mathbf{v}_{T,l}\rangle} \cdot e^{i\frac{2\pi}{\lambda d}\langle\mathbf{u}_{R,j},\mathbf{v}_{T,l}\rangle}\end{bmatrix}d\mathbf{u}_{R,i}\,d\mathbf{u}_{R,j}\,d\mathbf{v}_{T,k}\,d\mathbf{v}_{T,l}$$

$$= 2M^2 - M + \frac{M(M-1)^2}{\pi^4}\int_{\mathcal{D}}\int_{\mathcal{D}}\int_{\mathcal{D}}\int_{\mathcal{D}}\begin{bmatrix}e^{i\frac{2\pi}{\lambda d}\langle\mathbf{w},\mathbf{y}\rangle} \cdot e^{-i\frac{2\pi}{\lambda d}\langle\mathbf{x},\mathbf{y}\rangle}\\ \cdot e^{-i\frac{2\pi}{\lambda d}\langle\mathbf{w},\mathbf{z}\rangle} \cdot e^{i\frac{2\pi}{\lambda d}\langle\mathbf{x},\mathbf{z}\rangle}\end{bmatrix}d\mathbf{z}\,d\mathbf{y}\,d\mathbf{w}\,d\mathbf{x}$$

$$= 2M^2 - M + M(M^2 - 2M + 1)f\left(\frac{2|\mathcal{S}|}{\lambda d}\right),$$

where

$$f(c) = \frac{1}{\pi^4} \cdot \int_\mathcal{D} \int_\mathcal{D} \int_\mathcal{D} \int_\mathcal{D} \begin{bmatrix} e^{ic\langle \mathbf{w},\mathbf{y}\rangle} \cdot e^{-ic\langle \mathbf{x},\mathbf{y}\rangle} \\ \cdot e^{-ic\langle \mathbf{w},\mathbf{z}\rangle} \cdot e^{ic\langle \mathbf{x},\mathbf{z}\rangle} \end{bmatrix} d\mathbf{z}\, d\mathbf{y}\, d\mathbf{w}\, d\mathbf{x}.$$

Hence, finding an upper bound for $E\{|\upsilon|^4\}$ is equivalent to finding an upper bound for $f(c)$. Toward that end, we note that

$$f(c) = \frac{1}{\pi^3} \int_\mathcal{D} \int_\mathcal{D} \int_\mathcal{D} e^{ic\langle \mathbf{w},\mathbf{y}\rangle} \cdot e^{-ic\langle \mathbf{x},\mathbf{y}\rangle} \left[\frac{1}{\pi} \int_\mathcal{D} e^{-ic\langle \mathbf{w}-\mathbf{x},\mathbf{z}\rangle} d\mathbf{z}\right] \cdot d\mathbf{w}\, d\mathbf{x}\, d\mathbf{y}$$

$$= \frac{1}{\pi^3} \int_\mathcal{D} \int_\mathcal{D} \int_\mathcal{D} e^{ic\langle \mathbf{w},\mathbf{y}\rangle} \cdot e^{-ic\langle \mathbf{x},\mathbf{y}\rangle} \left[\frac{1}{\pi} \int_{r=0}^1 \int_{\theta=0}^{2\pi} e^{-ic\langle \mathbf{w}-\mathbf{x},(\cos\theta,\sin\theta)\rangle} r\, d\theta\, dr\right] \cdot d\mathbf{y}\, d\mathbf{w}\, d\mathbf{x}.$$

Letting $\mathbf{w}' = \mathbf{w} - \mathbf{x} = \|\mathbf{w}'\|\cos\phi$, we obtain

$$f(c) = \frac{1}{\pi^3} \int_\mathcal{D} \int_{\mathcal{D}-\mathbf{x}} \int_\mathcal{D} e^{ic\langle \mathbf{w}'+\mathbf{x},\mathbf{y}\rangle} \cdot e^{-ic\langle \mathbf{x},\mathbf{y}\rangle} \left[\frac{1}{\pi} \int_{r=0}^1 \int_{\theta=0}^{2\pi} e^{-icr\|\mathbf{w}'\|\cos(\theta-\phi)} r\, d\theta\, dr\right] \cdot d\mathbf{y}\, d\mathbf{w}'\, d\mathbf{x}$$

$$= \frac{1}{\pi^3} \int_\mathcal{D} \int_{\mathcal{D}-\mathbf{x}} \int_\mathcal{D} e^{ic\langle \mathbf{w}'+\mathbf{x},\mathbf{y}\rangle} \cdot e^{-ic\langle \mathbf{x},\mathbf{y}\rangle} \left[\frac{2}{\pi} \int_{r=0}^1 \int_{\theta=0}^{\pi} \cos(cr\|\mathbf{w}'\|\cos(\theta-\phi)) r\, d\theta\, dr\right] \cdot d\mathbf{y}\, d\mathbf{w}'\, d\mathbf{x}$$

$$= \frac{2}{\pi^3} \int_\mathcal{D} \int_{\mathcal{D}-\mathbf{x}} \int_\mathcal{D} e^{ic\langle \mathbf{w}'+\mathbf{x},\mathbf{y}\rangle} \cdot e^{-ic\langle \mathbf{x},\mathbf{y}\rangle} \left[\int_{r=0}^1 J_0(cr\|\mathbf{w}'\|) r\, dr\right] \cdot d\mathbf{y}\, d\mathbf{w}'\, d\mathbf{x}$$

$$= \frac{2}{\pi^2} \int_\mathcal{D} \int_{\mathcal{D}-\mathbf{x}} \left[\int_{r=0}^1 J_0(cr\|\mathbf{w}'\|) r\, dr\right] \left[\frac{1}{\pi} \int_\mathcal{D} e^{ic\langle \mathbf{w}',\mathbf{y}\rangle} d\mathbf{y}\right] d\mathbf{w}'\, d\mathbf{x}$$

$$= \frac{4}{\pi^2} \int_\mathcal{D} \int_{\mathcal{D}-\mathbf{x}} \left[\int_{r=0}^1 J_0(cr\|\mathbf{w}'\|) r\, dr\right] \left[\int_{\rho=0}^1 J_0(c\rho\|\mathbf{w}'\|) \rho\, d\rho\right] d\mathbf{w}'\, d\mathbf{x}$$

$$= \frac{4}{\pi^2 c^4 \|\mathbf{w}'\|^4} \int_\mathcal{D} \int_{\mathcal{D}-\mathbf{x}} \left[\int_{r=0}^{c\|\mathbf{w}'\|} J_0(r) r\, dr\right]^2 d\mathbf{w}'\, d\mathbf{x},$$

where $J_0(r)$ is the $0^{\text{th}}$-order Bessel function of the first kind. Noting that $|J_0(r)| \leq r^{-1/2}$ for all $r \geq 0$, we get

$$f(c) \leq \frac{4}{\pi^2 c^4 \|\mathbf{w}'\|^4} \cdot \int_\mathcal{D} \int_{\mathcal{D}-\mathbf{x}} \left[\int_{r=0}^{c\|\mathbf{w}'\|} r^{1/2}\, dr\right]^2 d\mathbf{w}'\, d\mathbf{x} = \frac{16}{\pi^2 9c} \cdot \int_\mathcal{D} \int_{\mathcal{D}-\mathbf{x}} \|\mathbf{w}'\|^{-1} d\mathbf{w}'\, d\mathbf{x} \leq \frac{64}{9\pi c}.$$

Hence,

$$E\{\tilde{\xi}_M(\tilde{\mathbf{H}})\} \geq \frac{M^3}{4} \cdot \log_2\left(1+\frac{\gamma g}{2M^3}\right)\frac{1}{E\{|\upsilon|^4\}} \geq \frac{\frac{M}{4}\log_2\left(1+\frac{\lambda g}{2M^2}\right)}{\left(2-\frac{1}{M}\right)+\frac{32}{9\pi}\left(M-2+\frac{1}{M}\right)\Big/\frac{|\mathcal{S}|}{\lambda d}},$$

as claimed. ∎